# Observation of intrinsic inverse spin Hall effect

Lalani K. Werake, Brian A. Ruzicka, and Hui Zhao*
*Department of Physics and Astronomy, The University of Kansas, Lawrence, Kansas 66045, USA*

We report observation of intrinsic inverse spin Hall effect in undoped GaAs multiple quantum wells with a sample temperature of 10 K. A transient ballistic pure spin current is injected by a pair of laser pulses through quantum interference. By time-resolving the dynamics of the pure spin current, the momentum relaxation time is deduced, which sets the lower limit of the scattering time between electrons and holes. The transverse charge current generated by the pure spin current via the inverse spin Hall effect is simultaneously resolved. We find that the charge current is generated well before the first electron-hole scattering event. Generation of the transverse current in the scattering-free ballistic transport regime provides unambiguous evidence for the intrinsic inverse spin Hall effect.

The spin Hall effect was originally predicted in 1971,[1] and was revisited more recently in a number of theoretical works.[2–5] In this effect, a charge current produces a transverse pure spin current due to the spin-orbit coupling. This effect provides an electrical method to generate spin currents, which is a fundamental tool for spintronics. The first experimental observations[6, 7] have stimulated extensive experimental[8–12] and theoretical[13–21] studies of this effect. Based on the same physics mechanism, a pure spin current can generate a transverse charge current. Such an inverse spin Hall effect has also been proposed[2] and experimentally observed.[10, 22, 23] It provides an electrical method to detect spin current, which is also important for spintronics.

Despite these extensive efforts, a fundamental question of the spin Hall effect is still open: what are the *mechanisms* of the effect. When originally proposed, the effect was based on spin-dependent scattering between electrons and impurities.[1–3] This is now referred to as an extrinsic spin Hall effect. The *intrinsic* spin Hall effect that does not rely on scattering has also been proposed.[4, 5, 24, 25] Most reported experimental observations were performed in systems composed of thermalized carriers and were attributed to the extrinsic effect. An experimental study of the intrinsic spin Hall effect is of fundamental importance in studies of the spin-orbit interaction. It is also crucial for nanoscale spintronic devices where the transport is dominated by ballistic processes with no scattering.

Previously, one of us, along with co-workers, has demonstrated that ballistic currents injected optically in undoped GaAs samples by a quantum interference and control technique can induce spin Hall currents.[10] Because of technical limitations, the dynamics of these currents could not be time-resolved. Since then, we have significantly improved the current detection techniques.[26, 27] This allows us to study the generation mechanisms of the spin Hall currents, and directly observe the intrinsic inverse spin Hall effect. In the experiments, we instantaneously injected a transient ballistic pure spin current in undoped GaAs multiple-quantum-well samples by a pair of femtosecond laser pulses through quantum interference. After injection, the pure spin current decays due to scattering. By time resolving the decay, we determined the scattering time. We simultaneously monitored the dynamics of the generated transverse charge current, and found that the charge current is established well before the average time of the first scattering event. Since the extrinsic effect induced by the spin-dependent scattering can be safely excluded in this process, the intrinsic effect is unambiguously observed.

We studied three undoped GaAs/Al$_{0.3}$Ga$_{0.7}$As multiple-quantum-well samples with different periods and quantum-well thicknesses that were grown along the [001]-direction. Samples A and B contains 40 periods of 7.4 and 10 nm quantum wells, respectively. Sample C has 20 periods of 14-nm quantum wells. In each sample, the thicknesses of the barriers and the quantum wells are equal. In the measurements, the samples are cooled to 10 K to reduce the phonon absorption rate. We will first present results from sample A, and then discuss results from the other two samples.

Figure 1 summarizes our experimental approach. The sample is simultaneously illuminated by two tightly focused laser pulses that copropagate along the $\hat{z}$-direction with angular frequencies $\omega$ and $2\omega$, respectively (waves in Fig. 1A). Electrons can be excited from the valence band to the conduction band by one-photon absorption of the 180-fs, 750-nm, and $\hat{y}$-polarized $2\omega$ pulse and two-photon absorption of the 100-fs, 1500-nm, and $\hat{x}$-polarized $\omega$ pulse (vertical arrows in Fig. 1B). Due to the interference of the two transition pathways, electrons with opposite spin orientations along $\hat{z}$ (orange and blue spheres) are injected to the conduction band with opposite crystal momenta.[28] In real space, spin-up and spin-down electrons are injected with opposite velocities along $\hat{x}$, resulting in a pure spin current along $\hat{x}$ (Fig. 1A). In the measurements, the [100] direction of the sample is along $\hat{x}$, however, the current injection process only weakly depends on the orientation.[28] The average velocity of each spin system is proportional to $\cos(\Delta\phi)$, where $\Delta\phi$ is the relative phase between the two transition amplitudes.[28] This allows control of the spin current density by the phase, without changing the carrier density. In our experiments, we choose $\Delta\phi = 0$ in order to get the maximum pure spin current injection.

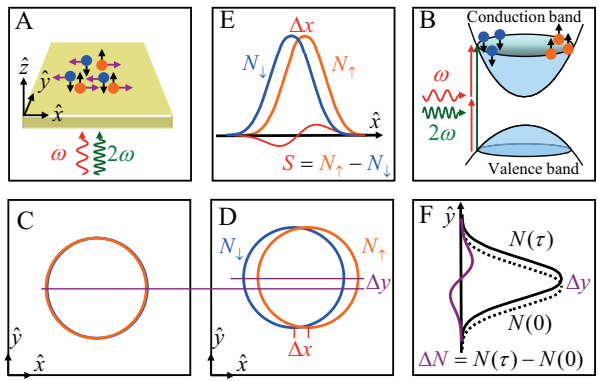

FIG. 1. The GaAs quantum-well sample is simultaneously illuminated by two laser pulses with angular frequencies of $\omega$ and $2\omega$ (waves in panels A and B). Quantum interference between one-photon absorption of the $2\omega$ pulse and the two-photon absorption of the $\omega$ pulse (vertical arrows in panel B) causes spin-up (orange spheres) and spin-down (blue spheres) electrons to be excited with opposite wavevectors, forming a pure spin current along $\hat{x}$. Initially, the two spin systems overlap in space (panel C). After a short period of time, they separate by a small distance $\Delta x$ (panel D), resulting in a spin density $S$ with a derivativelike profile along $\hat{x}$ (panel E). Due to the inverse spin Hall effect, both spin systems move along $\hat{y}$, causing the whole electron density profile ($N$) to move a distance $\Delta y$ (positive or negative) from the origin. The resulting electron accumulation $\Delta N$ has a derivativelike profile along $\hat{y}$ (panel F).

Upon injection, the spatial profiles of the spin-up and spin-down electrons overlap in space (Fig. 1C). After a short period of time, the two profiles separate by a distance $\Delta x$ (Fig. 1D). Because of the inverse spin Hall effect, both profiles are expected to move along $\hat{y}$, resulting in a displacement $\Delta y$ from the origin (Fig. 1D). Clearly, the time evolution of $\Delta x$ and $\Delta y$ reflects the dynamics of the pure spin current and the transverse charge current.

We measure $\Delta x$ and $\Delta y$ by using a differential transmission technique in a derivative detection geometry, as we have previous described.[10, 26] Figure 1E shows the spatial profiles of the densities of spin-up ($N_\uparrow$) and spin-down ($N_\downarrow$) electrons along $\hat{x}$. The spin density, $S \equiv N_\uparrow - N_\downarrow$, obviously has a derivativelike profile along $\hat{x}$. The height of this profile, $S^{\mathrm{MAX}}$, is related to the height and the width of the density profile of one spin system, $N_\uparrow^{\mathrm{MAX}}$ and $w$, by $S^{\mathrm{MAX}} = 1.4(N_\uparrow^{\mathrm{MAX}}/w)\Delta x$.[10] Similarly, Fig. 1F shows the spatial profiles of the total electron density, $N \equiv N_\uparrow + N_\downarrow$, along $\hat{y}$ before (dotted line) and after (solid line) the transport. The difference, $\Delta N \equiv N(\tau) - N(0)$, which can be viewed as transport-induced electron accumulation, has a derivativelike profile along $\hat{y}$, with a height related to $\Delta y$ in a similar fashion.[29] Therefore, as we have previously demonstrated,[10, 29, 30] we can determine $\Delta x$ and $\Delta y$ by measuring $S$ and $\Delta N$, even though these transport distances are much smaller than the direct spatial resolution of the system that is defined by the size of the laser spots.

We start our measurement by acquiring the profile of $N$ by scanning a probe spot, tuned to the heavy-hole excitonic resonance of sample A (790 nm), in the $x-y$ plane with a fixed probe delay of 0.5 ps, as shown in Fig. 2A. Since the two spin systems are injected with opposite average velocities along $\hat{x}$, the profiles of the two spin systems should separate along $\hat{x}$, resulting in a derivativelike $S$ profile, as illustrated in Fig. 1E. We measure $S$ as we scan the probe spot along the $\hat{x}$-direction with $y = 1$ $\mu$m (white horizontal line in Fig. 2A). The solid squares in Fig. 2B show the results, with the expected derivativelike profile. By comparing $S$ and $N$, we deduce that at $\tau = 0.5$ ps, $\Delta x = 17$ nm. The absolute signs of $S$ and $\Delta x$ are ambiguous. Because of the inverse spin Hall effect, we expect the profile of $N$ to move along $\hat{y}$, resulting in a nonzero $\Delta N$. We measure $\Delta N$ as we scan the probe spot along $\hat{y}$ with $x = 1$ $\mu$m (white vertical line in Fig. 2A). The solid circles in Fig. 2E show the results. The derivativelike profile of $\Delta N$ confirms the charge transport along $\hat{y}$. From this profile, we deduce $\Delta y = 0.3$ nm. Since the pump pulses do not inject a charge current with $\Delta\phi = 0$,[10, 28] the charge current along $\hat{y}$ is generated by the pure spin current along $\hat{x}$ via the inverse spin Hall effect. The simultaneously measured $S$ has a Gaussian-like profile along $\hat{y}$, as shown as the solid squares in Fig. 2E. This is because the two spin systems do not separate along $\hat{y}$ (see Fig. 1), and the transverse charge current is a *pure* charge current. Similarly, the simultaneously measured $\Delta N$ when the probe is scanned along $\hat{x}$ has a Gaussian-like profile along $\hat{x}$ (solid circles in Fig. 2C).

To resolve the dynamics of these currents, we fix the probe spot at $x = y = 1$ $\mu$m (white circle in Fig. 2A) and simultaneously measure $S$ and $\Delta N$ as we scan the probe delay. Figure 2F shows the $\Delta x$ (solid squares, left axis) and $\Delta y$ (solid circles, right axis) deduced from the measured $S$ and $\Delta N$ as a function of the probe delay. The increase of $\Delta x$ is caused by the spin transport, and is slowed down by the scattering. The line over the solid squares in Fig. 2F shows a fit to the data with a momentum relaxation time of each spin system of 0.45 ps.[30] The decrease of $\Delta y$ after the peak, i.e. the movement of electrons back towards origin, is caused by the space charge field of the holes, and has been previously observed when a charge current is optically injected.[29]

The momentum relaxation of each spin system is caused by scattering events, and therefore the relaxation time of the pure spin current is determined by the scattering time. Since the sample is undoped, impurity scattering is negligible in such an ultrafast process. (In fact, including the impurity scattering will not change our conclusion.) The scattering mechanisms contributing to the momentum relaxation include electron-hole scattering, phonon scattering, and scattering between electrons with opposite spin orientations. Scattering between electrons with the same spin orientation conserves the total momentum of each spin system, and therefore does not cause relaxation of the pure spin current. Among these scattering mechanisms, only the electron-hole scattering can possibly cause the transverse charge current via the extrinsic inverse spin Hall effect. Since the relaxation is caused by the three coexisting scattering mechanisms, the scattering time of the electron-hole scattering is at least (most likely longer than) 0.45 ps. Therefore, the charge current induced by the electron-hole scattering via the *extrinsic* inverse spin Hall effect can only be established on a time scale longer than 0.45 ps. However, the simultaneously measured $\Delta y$ reaches the maximum before 0.45 ps. Since the charge current density is proportional to the time-derivative of $\Delta y$, it reaches a peak even earlier than this. Since the charge current has been established well before the first scattering event, it cannot be due to the extrinsic inverse spin Hall effect.

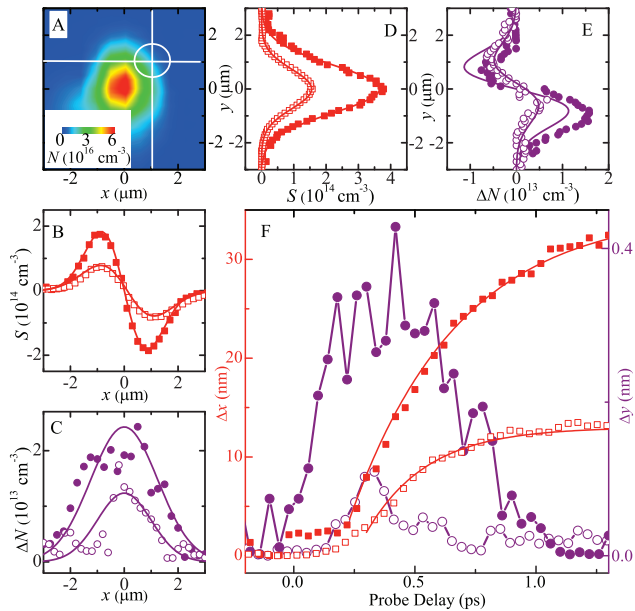

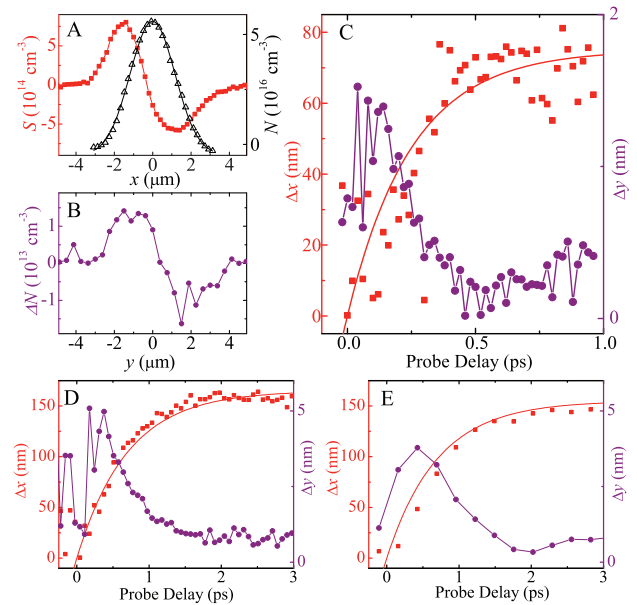

FIG. 2. Panel A shows the profile of the electron density ($N$) measured by scanning the probe spot in the x-y plane with a probe delay of 0.5 ps. The sample temperature is 10 K and the peak carrier density is $6 \times 10^{16}$ cm$^{-3}$. The spin density ($S$) and the electron accumulation ($\Delta N$) measured by scanning the probe spot along the horizontal line shown in panel A are plotted as the solid squares in panel B and the solid circles in panel C, respectively. Solid squares and solid circles in panels D and E show these two quantities measured by scanning the probe spot along the vertical line shown in panel A. Panel F shows the deduced $\Delta x$ (solid squares) and $\Delta y$ (solid circles) as a function of the probe delay. All the open symbols in panels B-F show corresponding results obtained with a higher peak carrier density of $2.4 \times 10^{17}$ cm$^{-3}$

FIG. 3. Panels A and B show measurements of sample B that correspond to Figs. 2B and 2E, respectively, with a fixed probe delay of 0.2 ps. Panels C and D corresponds to Fig 2F, but from samples B and C, respectively, both measured at $x = y = 1$ $\mu$m. Panel E is the same as panel D but measured at a different probe position ($x = 1, y = -1$ $\mu$m).

We repeated the experiment with a higher carrier density. The $S$ and $\Delta N$ measured at a fixed probe delay of 0.25 ps when the probe is scanned along the two white lines in Fig. 2A are plotted in Figs. 2B-2E as the open symbols. The time evolution of $\Delta x$ and $\Delta y$ is shown in Fig. 2F as the open symbols, as well. The momentum relaxation of each spin system is apparently faster, due to the increased scattering rate between electrons and holes. We deduce a momentum relaxation time of 0.25 ps. Similarly, the charge current is established before that time scale, confirming the intrinsic nature of the observed inverse spin Hall effect.

In order to verify that the observation is not specific to sample A, we have performed similar measurements in samples B and C. Here, no attempts were made to systematically investigate the influence of the sample structure, due to the complicated current injection and relaxation processes in samples with different structures. Panels A, B and C of Fig. 3 summarized results of sample B under the same experimental conditions except for a different carrier density of $9 \times 10^{16}$ cm$^{-3}$ and a different wavelength of the probe pulse of 795 nm (the heavy-hole exciton resonance of sample B). The transverse charge current is observed before the relaxation time of the pure spin current of 0.25 ps (solid line in Fig. 3C), that is consistent with results of sample A.

Figures 3D and 3E show the results of a similar mea-surement of sample C, measured at two probe locations. With a heavy-hole excitonic resonance of 800 nm, this sample allows us to tune the $\omega$ pulse to 1560 nm. Since the carriers are excited with an excess energy of 25 meV, below the optical phonon energy of GaAs (36 meV), optical phonon emission is not allowed. With a carrier density of $4\times10^{16}$ cm$^{-3}$, the relaxation time of the pure spin current increases to 0.6 ps (solid lines in Figs. 3D and 3E). Clearly, the charge current is generated much earlier than this time scale. The intrinsic nature of the inverse spin Hall effect in our experiment is confirmed again.

After the preliminary data of this work have been presented in a conference,[31] evidence of intrinsic spin Hall effect was reported.[32] In that work, the *ordinary* spin Hall effect in HgTe nanostructures was studied by steady-state electric measurements. However, the all-optical time-resolved technique allows us to study the *inverse* spin Hall effect in undoped GaAs quantum wells, and to provide direct dynamical evidence.

In conclusion, we have observed the intrinsic inverse spin Hall effect in undoped GaAs multiple quantum wells at 10 K. A transient ballistic pure spin current was injected by a pair of laser pulses through quantum interference. By time-resolving the dynamics of the pure spin current, we deduce a momentum relaxation time of each spin system, which sets the lower limit of the scattering time between electrons and holes. We simultaneously time-resolved the transverse charge current generated by the pure spin current via the inverse spin Hall effect. We found that the charge current is generated well before the first scattering event. Since we can safely exclude the extrinsic inverse spin Hall effect in this scattering-free ballistic regime,[4, 5, 24, 25, 33] we conclude that the observed inverse spin Hall effect is intrinsic.

We acknowledge John Prineas for providing us with high quality GaAs samples and E. Sherman for helpful discussions. This material is based upon work supported by the National Science Foundation of USA under Grant No. DMR-0954486.

* Corresponding author (Email: huizhao@ku.edu)
[1] M. I. D'yakonov and V. I. Perel, JETP Lett. **13**, 467 (1971).
[2] J. E. Hirsch, Phys. Rev. Lett. **83**, 1834 (1999).
[3] S. F. Zhang, Phys. Rev. Lett. **85**, 393 (2000).
[4] S. Murakami, N. Nagaosa, and S. C. Zhang, Science **301**, 1348 (2003).
[5] J. Sinova, D. Culcer, Q. Niu, N. A. Sinitsyn, T. Jungwirth, and A. H. MacDonald, Phys. Rev. Lett. **92**, 126603 (2004).
[6] Y. K. Kato, R. C. Myers, A. C. Gossard, and D. D. Awschalom, Science **306**, 1910 (2004).
[7] J. Wunderlich, B. Kaestner, J. Sinova, and T. Jungwirth, Phys. Rev. Lett. **94**, 047204 (2005).
[8] V. Sih, R. C. Myers, Y. K. Kato, W. H. Lau, A. C. Gossard, and D. D. Awschalom, Nat. Phys. **1**, 31 (2005).
[9] N. P. Stern, S. Ghosh, G. Xiang, M. Zhu, N. Samarth, and D. D. Awschalom, Phys. Rev. Lett. **97**, 126603 (2006).
[10] H. Zhao, E. J. Loren, H. M. van Driel, and A. L. Smirl, Phys. Rev. Lett. **96**, 246601 (2006).
[11] N. P. Stern, D. W. Steuerman, S. Mack, A. C. Gossard, and D. D. Awschalom, Nat. Phys. **4**, 843 (2008).
[12] E. S. Garlid, Q. O. Hu, M. K. Chan, C. J. Palmstrøm, and P. A. Crowell, Phys. Rev. Lett. **105**, 156602 (2010).
[13] E. G. Mishchenko, A. V. Shytov, and B. I. Halperin, Phys. Rev. Lett. **93**, 226602 (2004).
[14] L. Sheng, D. N. Sheng, and C. S. Ting, Phys. Rev. Lett. **94**, 016602 (2005).
[15] C. L. Kane and E. J. Mele, Phys. Rev. Lett. **95**, 226801 (2005).
[16] S. Q. Shen, M. Ma, X. C. Xie, and F. C. Zhang, Phys. Rev. Lett. **92**, 256603 (2004).
[17] S. Zhang and Z. Yang, Phys. Rev. Lett. **94**, 066602 (2005).
[18] B. K. Nikolic, S. Souma, L. P. Zarbo, and J. Sinova, Phys. Rev. Lett. **95**, 046601 (2005).
[19] B. A. Bernevig and S. C. Zhang, Phys. Rev. Lett. **95**, 016801 (2005).
[20] H. Engel, B. I. Halperin, and E. I. Rashba, Phys. Rev. Lett. **95**, 166605 (2005).
[21] S. Murakami, N. Nagaosa, and S. C. Zhang, Phys. Rev. Lett. **93**, 156804 (2004).
[22] S. O. Valenzuela and M. Tinkham, Nature **442**, 176 (2006).
[23] T. Kimura, Y. Otani, T. Sato, S. Takahashi, and S. Maekawa, Phys. Rev. Lett. **98**, 156601 (2007).
[24] J. Schliemann and D. Loss, Phys. Rev. B **69**, 165315 (2004).
[25] J. Inoue, G. E. W. Bauer, and L. W. Molenkamp, Phys. Rev. B **70**, 041303 (2004).
[26] B. A. Ruzicka, K. Higley, L. K. Werake, and H. Zhao, Phys. Rev. B **78**, 045314 (2008).
[27] H. Zhao and A. L. Smirl, Appl. Phys. Lett. **97**, 212106 (2010).
[28] R. D. R. Bhat and J. E. Sipe, Phys. Rev. Lett. **85**, 5432 (2000).
[29] H. Zhao, E. J. Loren, A. L. Smirl, and H. M. van Driel, J. Appl. Phys. **103**, 053510 (2008).
[30] H. Zhao, A. L. Smirl, and H. M. van Driel, Phys. Rev. B **75**, 075305 (2007).
[31] L. K. Werake, B. A. Ruzicka, and H. Zhao, Quantum Electronics and Laser Science Conference (QELS) **May 19**, Talk QWC3 (2010).
[32] C. Brüne, A. Roth, E. G. Novik, M. König, H. Buhmann, E. M. Hankiewicz, W. Hanke, J. Sinova, and L. W. Molenkamp, Nat. Phys. **6**, 448 (2010).
[33] E. Y. Sherman, A. Najmaie, H. M. van Driel, A. L. Smirl, and J. E. Sipe, Solid State Commun. **139**, 439 (2006).